\documentclass[runningheads]{llncs}
\usepackage{graphicx}
\usepackage{booktabs}
\usepackage{amssymb}
\usepackage{amsmath}
\usepackage{cleveref}
\usepackage{makecell}
\usepackage{multirow}
\usepackage{url}
\begin{document}
%
\title{The Influence of Dataset Partitioning on Dysfluency Detection Systems}
%
\author{Sebastian P. Bayerl\inst{1} \and
 Dominik Wagner\inst{1} \and
 Elmar Nöth\inst{2} \and
 Tobias Bocklet\inst{2} \and
Korbinian Riedhammer \inst{1}}
\authorrunning{Sebastian P. Bayerl et al.}

\institute{Technische Hochschule Nürnberg Georg Simon Ohm \and
Friedrich Alexander Universität Erlangen-Nürnberg\\
\email{sebastian.bayerl@ieee.org}\\
}
%
%
\maketitle              
\begin{abstract}
This paper empirically investigates the influence of different
data splits and splitting strategies on the performance of dysfluency detection systems. For this, we perform experiments using \textit{wav2vec 2.0} models with a classification head as well as support vector machines (SVM) in conjunction with the features extracted from the \textit{wav2vec 2.0} model to detect dysfluencies. We train and evaluate the systems with different non-speaker-exclusive and speaker-exclusive splits of the Stuttering Events in Podcasts (SEP-28k) dataset to shed some light on the variability of results w.r.t. to the partition method used. Furthermore, we show that the SEP-28k dataset is dominated by only a few speakers, making it difficult to
evaluate. To remedy this problem, we created SEP-28k-Extended (SEP-28k-E), containing semi-automatically generated speaker and gender information for the SEP-28k corpus, and suggest different data splits, each
useful for evaluating other aspects of methods for dysfluency detection.
\end{abstract}
\section{Introduction}
Stuttering is a speech disorder that negatively affects a person's ability to communicate.
Detecting if speech is dysfluent has implications for stuttering self-help applications, monitoring of stuttering behaviour, therapy applications, and enabling speech recognition systems
to be more inclusive. 
Stuttering is highly individual and has a huge inter and intra-person variance w.r.t. the occurrence of symptoms, which among other things, depends on psychological factors, the communication situation, or the linguistic complexity of an utterance \cite{ellis_handbook_2009}.
Those properties make the detection of stuttering a very hard problem. 

Data scarcity has always been a problem for research on pathological speech in general, even more so with the rise of neural networks, which typically need large amounts of labeled training data. 
Datasets containing pathological speech are often small, inconsistently labeled, and rarely publicly available. 
The same was true for stuttering in the past, but recent efforts by various research groups are a step towards resolving the problem of data scarcity and being able to focus on detection methods. 

Kourkounakis et al. released LibriStutter, a dataset containing artificially generated stuttered speech with word-level stuttering annotations \cite{kourkounakis_libristutter_2021}. 
Stuttering Events in Podcasts (SEP-28k) is a large corpus consisting of some 28000 clips ($\sim$23 hours) annotated with five stuttering event types \cite{lea_sep-28k_2021}.
The authors also released a relabeled version of the \emph{adults who stutter} subset of FluencyBank using the same annotation scheme, adding another 4000 clips \cite{bernstein_ratner_fluency_2018,lea_sep-28k_2021}.
Using a similar labeling approach, Bayerl et al. introduced a therapy-centric dataset of stuttered speech containing data taken from stuttering therapy with an additional label for modified speech, indicating the use of speech technique as it is taught in stuttering therapy\cite{bayerl_ksof_2022}.

It is good practice to use speaker-exclusive data partitioning when training and testing machine learning systems on pathological speech and paralinguistic tasks (e.g., splits in \cite{schuller_computational_2014,schuller_interspeech_2016,schuller_interspeech_2018}).
This data partitioning method prevents that systems learn speaker-specific traits instead of the targeted behaviour. 
Even when using this data partitioning method, there are many possible combinations of speakers. 
Ideally, the distributions of classes, the speakers' gender, and age are similar among the train, test, and development sets. 
If this cannot be guaranteed, $N$-fold cross-validation is a possible way of partitioning the data but will increase the number of experiments necessary $N$ times.
Even with cross-validation, speakers in the train and development partition of a fold should not appear in the test partition of the same fold.
To be able to compare the performance of different machine learning systems on such a problem, it must be ensured that the data-partitioning can be reproduced \cite{schuller_computational_2014}.
If this is not the case, it is impossible to compare different methods' results.
Results reported on unknown, arbitrary splits are not representative and are of little use to other researchers. 

In the initial release of SEP-28k, the authors suggest several baseline systems and hint at their data partitioning \cite{lea_sep-28k_2021}, but unfortunately did not include the splits. 
There is little metadata available besides the podcast name and episode number the clips were taken from, making it hard to create an ideal speaker-exclusive data split.
Using this information only allows for leave-one-podcast out validation, to not include speakers in the test fold that were in the training or development partition.
This metadata is available for the relabeled FluencyBank portion of their data, but unfortunately, the baseline results were not reported with the split used.
  
In this paper, we evaluate the influence of different non-speaker and speaker-exclusive data partitioning methods using a frozen wav2vec 2.0 model with a classification head and use the same features in conjunction with Support Vector Machine (SVM) classifiers. 
We provide additional insights into the composition of the SEP-28k dataset and describe a process to generate per-episode speaker labels based on manually collected meta-data and ECAPA-TDNN embeddings. 
The additional meta-data is made available to the scientific community in an updated release called SEP-28k-Extended. 
Furthermore, we report baseline results for newly created splits which consider the retrieved metadata.

\section{Data}\label{sc:data}
In this paper, we use and extend data from the SEP-28k dataset. 
The dataset consists of 385 podcast episodes taken from eight podcasts revolving around the topic of stuttering. 
It contains $\sim$28000 3sec long clips extracted from the episodes and labeled with five dysfluency types; blocks, sound repetitions, interjections, word repetitions, and prolongations \cite{lea_sep-28k_2021}.
The labels are non-exclusive, meaning a clip can be labeled as belonging to more than one class. 
The initial release of the dataset also contains $\sim$ 4000 clips extracted from the \emph{adults who stutter} portion of FluencyBank, that were labeled similarly \cite{bernstein_ratner_fluency_2018,lea_sep-28k_2021}.

For this paper, we researched missing metadata from all 385 episode descriptions to extract the number of speakers per episode, i.e., the maximum number of speakers that can be expected in the extracted clips of each episode. 
A closer examination of the label distribution (see \Cref{tbl:distribution}) of each podcast and the statistics for the whole dataset reveals a large imbalance w.r.t. the distribution of labels between the individual podcasts and the number of clips in the total dataset. 
A substantial share of clips was extracted from the podcast Women Who Stutter (WWS), about 33\% of total clips, followed by three other podcasts that each add $\geq$ 10\% of total clips.
Analysis of the retrieved metadata reveals that the He Stutters (HS) podcast is hosted by the same women as WWS, interviewing men instead of women. 
Aggregating those two podcasts shows an even greater imbalance, which means that one female speaker could be in up to 46\% of total clips.  
This has a potentially negative effect on the generalisation ability of detection systems trained on this data, even though the dataset is rather big. 
For experiments in this paper, we will therefore treat the two podcasts as one.

\begin{table}[tb]
\caption{Distribution of stuttering-related labels in SEP-28k per podcast, total number of clips, and share of the complete dataset.}\label{tbl:distribution}

\centering
\begin{tabular}{l|c|c|c|c|c|c|c|c||c}
\toprule
             &  HVSA &   ~ISW~ &   ~MSL~ &    ~SV~ &    ~ST~ &   ~SIC~ &   ~WWS~ &    ~HS~ &  ~total~\\   
\midrule                                                                                              
\textbf{Block} & 12.23 & 14.71 & 10.09 & 15.16 & 10.78 &  8.80 & 13.59 & 11.45 &  12.10                \\   
\textbf{Interjection} & 33.70 & 20.00 & 13.51 &  6.76 & 23.72 & 22.43 & 21.96 & 26.22 &  21.04                \\   
\textbf{Prolongation} & 11.96 & 11.84 &  7.87 & 14.99 &  8.33 &  7.87 &  9.89 & 12.13 &  10.61                \\   
\textbf{Sound repetition} & 11.41 & 29.08 &  6.71 &  1.60 &  8.04 &  5.18 &  8.07 & 12.40 &  10.31                \\   
\textbf{Word repetition } &  6.66 &  6.32 &  6.46 &  1.95 & 11.45 &  8.35 & 11.38 & 13.90 &   8.31                \\   
\textbf{No stuttered words} & 51.49 & 42.41 & 64.77 & 62.52 & 59.46 & 65.94 & 53.34 & 48.72 &  56.08                \\   
\midrule                                                                                                
\textbf{Total \#}  & 736   & 870   & 2339  & 2308  & 5064  & 4013  & 9163  & 3684  & 28177                 \\   
\textbf{\% of total}  & 2.61  & 3.09  & 8.30  & 8.19  & 17.97 & 14.24 & 32.52 & 13.07 & 100.00                \\   
\bottomrule
\end{tabular}

\end{table}

\section{Methods}

\subsection{Classification Experiments}
We chose a simple experimental design based on the \textit{wav2vec 2.0} model to evaluate the influence of data partitioning on the detection of atypical speech patterns. 
W2V2 features have shown robust performance in several speech tasks, such as speech recognition, speech emotion recognition, and mispronunciation detection \cite{baevski_wav2vec_2020,pepino_emotion_2021,xu_explore_2021}.

We use a W2V2 model that was unsupervised pre-trained on 960 hours of unlabeled speech data from the LibriSpeech corpus \cite{panayotov_librispeech_2015} and later fine-tuned for automatic speech recognition (ASR) on the transcripts of the same data. 
The weights for the model were published by \cite{baevski_wav2vec_2020}.
The model yields different hidden representations after each of the models' twelve transformer blocks.
Depending on the location in the processing hierarchy, the model has encoded different information into the representations, with embeddings from lower layers having basic speech information encoded and higher layers encoding information closer to phonemic information.
The W2V2 model uses the self-attention mechanism that helps the model focus on relevant parts of the input sequence, w.r.t. the learning objective \cite{vaswani_attention_2017}.
The hidden representations have information about their relationship to other vectors in the extraction context encoded. 
The model takes the raw wave-form audio as its inputs and yields 768-dimensional speech representations for roughly every 0.02s of audio, yielding 149 vectors for every 3s long clip in the dataset. 

For our experiments, we used frozen W2V2 models with a classification head equivalent to the implementation from the Transformers library \cite{wolf_transformers_2020}.
The classification head consists of a mean-pooling operation, pooling model outputs over the time dimension, yielding a single 768-dimensional vector for every audio clip. 
The pooling operation is followed by a 256-dimensional dense projection layer and a classification layer.

The same frozen model is also used to extract the contextual W2V2 embeddings as input features for training Support Vector Machine (SVM) classifiers, 
as they allow quick experimentation and can learn from only a few samples.
We extract W2V2 features for every three-second long clip and, similar to the mean-pooling operation in the classification head described previously, take the mean over the time dimension, yielding one 768-dimensional vector for every three-second long audio clip.  

\subsection{ECAPA-TDNN}
The Emphasized Channel Attention, Propagation, and Aggregation - Time Delay Neural Network (ECAPA-TDNN) architecture builds on the x-vector architecture for speaker identification and proposes several enhancements \cite{Desplanques2020ecapa,snyder_x-vectors_2018}. 
The two main modifications are 1-dimensional Res2Net \cite{gao2021res2net} modules with skip connections and squeeze-and-excitation (SE) \cite{jie2018squeeze} blocks to capture channel interdependencies. 
Furthermore, features are aggregated and propagated across multiple layers. 

We use the ECAPA-TDNN implementation from \cite{speechbrain}. 
The model was trained on the VoxCeleb dataset \cite{Nagrani17}.
The training data is augmented with additive noises from the MUSAN corpus \cite{musan2015} and reverberation using a collection of room impulse responses \cite{rirs2017}. 
It uses 80-dimensional Mel Frequency Cepstral Coefficients (MFCC) with a frame width of 25 ms and a frame-shift of 10 ms as its' inputs. 
Additionally, the data is speed-perturbed at 95\% and 105\% of the normal utterance speed, and the SpecAugment \cite{specaugment2019} method is applied in the time domain. 

\subsection{Metadata retrieval}
We use the ECAPA embeddings to automatically generate speaker labels for each of the clips in the SEP-28k dataset, allowing more granular speaker exclusive splits than on the podcast.  
The assignment of a speaker to individual audio clips is accomplished in an unsupervised manner using K-Means clustering and silhouette analysis \cite{ROUSSEEUW198753}. 

We employ silhouette analysis to assess the distance of separation between clusters without ground truth labels being available. 
The silhouette coefficient $s$ for an individual data point $x$ is given by $s_x = (b-a)(\text{max}(a, b))^{-1}$.
The variable $a$ represents the average distance between the sample and all other points in the same cluster. 
The variable $b$ is the mean distance between the sample and all other points in the nearest cluster.

The measure has a value range of $[-1, 1]$. 
Silhouette coefficients close to $+1$ indicate that the sample is far away from neighboring clusters and therefore likely assigned correctly. 
A value of $0$ indicates that the sample is close to the decision boundary between two neighboring clusters and negative values indicate that those samples might have been assigned to the wrong cluster.

Silhouette analysis can also be used to determine the optimal number of clusters. 
For a set of cluster values $k \in \lbrace m, \ldots, n \rbrace \subset \mathbb{N}_{>0}$, the optimal number of clusters $k_{opt}$ can be chosen as the one that maximizes the global average silhouette coefficient. 
We employ this method to episodes where the number of guests could not be determined.

The process of generating speaker labels and determining podcast hosts starts with the extraction of high-dimensional ($\mathbb{R}^{192}$) speaker embeddings from the trained ECAPA-TDNN model. 
We collect all embeddings belonging to one podcast, preprocess the embeddings, and subsequently cluster them using the K-Means algorithm. 
The preprocessing pipeline involves removing the mean, scaling to unit variance, and dimensionality reduction using Principal Component Analysis (PCA). 
We reduce the embedding dimensionality to $\mathbb{R}^{4}$, since it led to more robust distance computations, while the principal components still explained $\sim$33\% of total variance. 

We assume that the largest cluster for each podcast belongs to the podcast's host. 
This assumption is reasonable as most podcasts have the same host, who speaks across multiple episodes, while the guests and co-hosts vary across different episodes.
After preprocessing and clustering, we obtain the cluster centroids and select the one belonging to the largest cluster.
The centroid of the host cluster serves as a prototype vector, against which other clip-level representations are compared to determine whether they belong to the host.

We also fit individual cluster models for each podcast episode. 
The preprocessing and clustering steps are equivalent to the host centroid creation. 
In this case, the K-Means algorithm is applied to embeddings representing clips from a the same podcast episode. 
The resulting cluster labels serve as labels for the different speakers in an episode. 
The cluster centroids obtained on the episode-level are then compared to the global host centroids to determine which cluster label belongs to the host speaker. 
We compute the cosine distance between the global host centroid of the podcast and each centroid representing a cluster in a specific podcast episode. 
The smallest cosine distance indicates which cluster label is the best candidate for the host speaker.

\subsection{Quality Criteria} 
We analyze the quality of the automatically generated speaker labels based on the sample-specific silhouette score, average per-episode silhouette score, and the variance ratio criterion  \cite{calinski74}.
The measures and their respective per clip and episode values are included in the meta-data published with this work. 
Depending on the desired quality level of the speaker labels, it is possible to exclude clips for stricter evaluation.

Higher sample-specific silhouette scores indicate that the speaker label is more likely to be assigned correctly. 
The average silhouette score can be used to assess the assignment quality within full podcast episodes. 
The absence of clusters where all sample-specific scores are below the average silhouette score for the cluster, can be a quality indicator. 
For instance, clips belonging to an episode could be excluded if not all $k$ clusters contain at least one sample with an above-average silhouette score. 


The variance ratio criterion defined in \cite{calinski74} measures how well-defined clusters are. 
For a dataset $D$ with $n_D$ elements, which has been divided into $k$ clusters, the variance ratio criterion $v$ is defined as the ratio of the between-cluster dispersion and the within-cluster dispersion for all clusters.
A threshold could be set to filter out episodes with less well-defined clusters.
The score is higher when clusters are dense and well separated.

\begin{table}[tb]
    \centering
    \caption{Statistics for cluster quality measures aggregated across all podcast episodes in SEP-28k containing more than one speaker.}
    \label{tab:episodestats}
\begin{tabular}{llcccc}
\toprule
Model                  &      Measure                    &  ~~Mean &   ~~Standard Deviation &   ~~Minimum &    ~~Maximum \\
\midrule
& ~~Silhouette Score             &  ~~0.46 &  ~~0.09  &  ~~0.17 &   ~~0.68 \\
K-Means & ~~Variance Ratio   & ~~69.50 & ~~83.06 &  ~~8.52 &    ~~406.21 \\
& ~~Cosine Distance              &   ~~0.34 &  ~~ 0.46 &   ~~0.00 &    ~~1.96 \\
\bottomrule
\end{tabular}
\end{table}

We chose to use K-Means for clustering instead of Gaussian Mixture Models (GMM), as we found the slightly better quality criteria (e.g., average silhouette score of 0.46 vs. 0.44). 
The impact of setting different thresholds for silhouette score and variance ratio is illustrated in \Cref{tab:episodefulfilled}. 
The two columns under \textit{Combined Fulfilled} show the number of podcast episodes in which \textit{all three} quality criteria are fulfilled, and their share of the total number of episodes. 
The three criteria are also shown individually under \textit{Standalone Fulfilled}. 
The first criterion measures the number of episodes in which at least one sample silhouette score in each of the episodes' clusters is above the average silhouette score. 
The criterion is independent of the  thresholds of the other criteria and   therefore remains constant.  
The second and third criteria measure the number of episodes where the average silhouette score and the variance ratio are above the threshold displayed below the \textit{Threshold} columns. 

\begin{table}[tb]
    \centering
    \caption{Impact of varying silhouette score and variance ratio thresholds on the quality of speaker labels obtained using K-Means clustering. 
    Thirty-three episodes were excluded from the analysis since they contained only one speaker. The numbers below indicate the number of episodes meeting the specified quality criterion thresholds.}
    \label{tab:episodefulfilled}
\begin{tabular}{lc|ccc|cc}
\toprule
\multicolumn{2}{c}{\textbf{Threshold} }                &                     \multicolumn{3}{c}{\textbf{Standalone Fulfilled}}                & \multicolumn{2}{c}{\textbf{Combined Fulfilled}}\\
 \makecell{Silhouette \\ Score}  &  \makecell{Variance\\ Ratio}  & \makecell{1.\\All Above\\Average} &  \makecell{2.\\Silhouette \\ Score}  & \makecell{3.\\Variance\\Ratio} &   Episodes & \makecell{\% of \\ Total} \\
\midrule
                 0.20 &                 10 &        278 &            332 & 331 &               270 &      76.7\% \\
                 0.30 &                 20 &        278 &            317 & 281 &                234 &      66.5\% \\
                 0.40 &                 30 &        278 &            272 & 197 &             166 &      47.2\% \\
                 0.50 &                 40 &        278 &            123 & 135 &               72 &      20.5\% \\
\bottomrule
\end{tabular}
\end{table}

\section{SEP-28k-Extended}
The extended data set created for this paper, SEP-28k-Extended (SEP-28k-E), is available online.\footnote{\protect \url{https://github.com/th-nuernberg/ml-stuttering-events-dataset-extended}}
It contains information about the gender and number of speakers to expect in the clips taken from each episode, and a label, identifying the podcast hosts. 
The dataset repository contains the original content of SEP-28k with the additional metadata and instructions for using the data.

\subsection{Speaker Imbalance}
Across all 385 podcast episodes, there are 42 (11\%) episodes in which the estimated speaking time of a single speaker (most likely the host) is above 90\% and 92 (24\%) episodes in which the estimated speaking time is above 80\%. 
This indicates that the dataset is dominated by a few speakers, which possibly has a detrimental influence on the generalisation ability and validity of evaluation results.
\Cref{tab:spkdom} shows the share of clips belonging to various podcast hosts relative to the total number of clips in a podcast. 
The StrongVoices (SV) podcast has two hosts, which makes the automatic assignment of a cluster label to the host speaker more difficult, so we excluded it from the analysis.
There are about 500 unique speakers in the dataset, but \Cref{tab:spkdom} displays the strong dominance of only four speakers, which are in 59\% of all clips.
This makes it very difficult to split the dataset so that it does not introduce a strong bias w.r.t. the dominant speakers.
\begin{table}[t]
    \centering
    \caption{Shares of clips that belong to podcast hosts in the  SEP-28k dataset.}
    \vspace{-2mm}
    \label{tab:spkdom}
    \begin{tabular}{llcccc}
\toprule
         Podcast &          ~~Host &    ~Clips~ & Host Share & ~\% of total clips~ &  Cumulative \\
\midrule
 WomenWhoStutter &  ~~Pamela Mertz &   9163 &      69.2\% &             22.63\%&       22.63\% \\
      HeStutters &  ~~Pamela Mertz &   3684 &      71.5\% &              9.40\%&       32.02\% \\
     StutterTalk & ~~Peter Reitzes &   5064 &      63.2\% &             11.42\%&       43.45\% \\
StutteringIsCool & ~~Daniele Rossi &   3853 &      66.6\% &              9.16\%&       52.60\% \\
MyStutteringLife &    ~~Pedro Peña &   2339 &      79.4\% &              6.63\%&       59.23\% \\
            HVSA &     ~~TJ Travor &    736 &      53.0\% &              1.39\%&       60.62\% \\
  IStutterSoWhat &  ~~Evan Sherman &    870 &      43.8\% &              1.36\%&       61.98\% \\
\bottomrule

\end{tabular}
\end{table}
\subsection{Data partitioning considering metadata}\label{ss:partitions}
This section briefly describes four different speaker-exclusive dataset splits, that were created considering the peculiarities of SEP-28k.
Each split has its purpose and tests another facet of a detection method.
The priorities for creating test, development, and training set, in order, where speaker-exclusiveness, label distribution, and gender distribution; statistics for the newly created splits can be viewed in \Cref{tbl:sep_e_dist}.

The \textbf{SEP-12k} split consists of about 12,000 clips taken from the original dataset.
These are all clips that are not associated with the top-four dominant speakers. 
We suggest evaluating the split using five-fold cross-validation without overlapping speakers between the folds.
The use of this split tests a method's ability to use many speakers with only few samples for training while also evaluating the method on many unseen speakers.

The \textbf{SEP-28k-E} split is partitioned into training, development, and test.
The training partition contains only clips belonging to the top-four dominant speakers.
The split can be used to test a method's ability to learn from many examples provided by few speakers. 

\begin{table}[]
    \caption{Composition of train, test and development set of SEP-28k-E and SEP-12k}
    \vspace{-2mm}
    \label{tbl:sep_e_dist}
    \centering
    \begin{tabular}{l|c|c|c|c}
\toprule
                            & SEP-28k-E train &     SEP-28k-E dev &    SEP-28k-E test  &      SEP-12k \\
\midrule                       
\textbf{Block}              & 11.57 \% &    12.84 \% &   12.01 \% &        12.48 \% \\
\textbf{Interjection}       & 22.94 \% &    18.79 \% &   19.51 \% &        19.10 \% \\
\textbf{Prolongation}       &  9.87 \% &    10.07 \% &   10.13 \% &        10.15 \% \\
\textbf{Sound repetition}   &  8.13 \% &    10.40 \% &    6.69 \% &         8.57 \% \\
\textbf{Word repetitions}   &  9.98 \% &     8.79 \% &   10.48 \% &         9.67 \% \\
\textbf{No stuttered words} & 56.92 \% &    55.78 \% &   58.15 \% &        56.81 \% \\
\midrule
\textbf{Total \#}                 & 15213 &     6402 &    6562 &        12804 \\

\bottomrule
\end{tabular}

\end{table}

The \textbf{SEP-28k-T} and \textbf{SEP-28k-D} splits are similar to each other as well as to SEP-28k-E.
They can be used to evaluate a method's capability to train on relatively few samples by many different speakers and  review its' performance on only a few speakers with many samples. 
The test partition consists of the four dominant speakers, and the development and training set are an equal size split of the remaining clips.
SEP-28k-D uses the same partitioning as SEP-28k-T but switches the test and development partitions. 

\section{Experiments}
The experiments described in this section are formulated as binary classification tasks that vary mostly w.r.t. the data partitioning used. 
To evaluate the influence of different data partitioning strategies on classification results, we performed experiments using a classification head on top of a frozen W2V2 feature extractor, and  SVM classifiers with radial basis function (rbf) kernels on seven different data-partitioning strategies; a leave-one-podcast out strategy, speaker agnostic five- and ten-fold cross-validation, three different  balanced, speaker separated Train-Validation-Test splits considering the additional meta-data from SEP-28k-E, and five-fold speaker-separated cross-validation on the SEP-12k subset. 
All results reported are F1-scores for the dysfluency classes, which is the harmonic mean of precision and recall and was also used for evaluation of the baseline systems by the original authors of SEP-28k \cite{lea_sep-28k_2021}.

The leave-one-podcast out strategy uses data from all podcasts for training and validation and uses data from the left-out podcast for testing. 
Both podcasts, He Stutters and Women Who Stutter, have the same host; we, therefore treat them as one podcast labeled as HeShe in the classification experiments.  
Both cross-validation splits are performed completely agnostic to the speaker and podcast label.
All cross-validation experiments were performed five times. 
We report the mean and the respective standard deviation of these results.

The classification head on top of the W2V2 model is equivalent to the implementation from the Transformers library \cite{wolf_transformers_2020}.
The training uses a single weighted cross-entropy loss term with an initial learning rate of 0.001, a batch size of 200, training for up to 200 epochs using the adam optimizer, with early stopping if the development loss is not decreasing for 10 epochs. 
The optimal position of the classification head after W2V2 transformer layer $L$ was determined from $L \in \{1, 2, \ldots, 12\}$ using cross-validation on the respective development partition. 

The optimal hyperparameters and input features for the SVM classifiers were determined using grid search in five-fold cross-validation on the respective development sets described in \cref{sc:data}.
The kernel parameter $\gamma$ was selected from the set
$\gamma \in \{10^{-k} \mid k = 1, \ldots, 4 \} \subset \mathbb{R}_{>0}$,
the regularisation parameter $C$ was selected from
$C \in \{ 1, 10, 100 \} \subset \mathbb{N}_{>0}$, and the W2V2 extraction layer $L$ was selected from $L \in \{1, 2, \ldots, 12\}$.

\begin{table}[t]
\caption{Results (F1-Scores) for experiments with leave-one-podcast out evaluation. Column headers are indicating the podcast name used as test-set.}\label{tbl:results}

\vspace{-2mm}
\centering
\begin{tabular}{r|r|c|c|c|c|c|c|c||c}
\toprule
&            &  HVSA & ISW   & MSL  & SV   &  ST  & SIC  & HeShe &  LOPO average     \\
\midrule                                                                        
\multirow{5}{25pt}{SVM}& \textbf{Blocks} & 0.34  & 0.37  & 0.32 & 0.40 & 0.33 & 0.29 & 0.36  &   0.34 (0.03)   \\
& \textbf{Interjections} & 0.73  & 0.68  & 0.62 & 0.48 & 0.73 & 0.71 & 0.70  &   0.66 (0.08)   \\
& \textbf{Prolongations} & 0.53  & 0.50  & 0.46 & 0.54 & 0.45 & 0.44 & 0.44  &   0.48 (0.04)   \\
& \textbf{Sound repetitions} & 0.39  & 0.70  & 0.38 & 0.19 & 0.45 & 0.36 & 0.41  &   0.41 (0.14)   \\
& \textbf{Word repetitions} & 0.34  & 0.35  & 0.43 & 0.31 & 0.50 & 0.42 & 0.49  &   0.41 (0.07)   \\
\midrule                                                                      
\multirow{5}{25pt}{NN}& \textbf{Blocks} & 0.19  & 0.21  & 0.16 & 0.23 & 0.19 & 0.15 & 0.17  & 0.19 (0.03)   \\
& \textbf{Interjections} & 0.65  & 0.63  & 0.58 & 0.44 & 0.71 & 0.68 & 0.68  & 0.62 (0.09)   \\
& \textbf{Prolongations} & 0.42  & 0.37  & 0.44 & 0.45 & 0.31 & 0.40 & 0.39  & 0.40 (0.05)   \\
& \textbf{Sound Repetitions} & 0.27  & 0.62  & 0.40 & 0.14 & 0.38 & 0.36 & 0.37  & 0.36 (0.14)   \\
& \textbf{Word Repetitions} & 0.25  & 0.26  & 0.42 & 0.34 & 0.42 & 0.37 & 0.40  & 0.35 (0.07)   \\
\bottomrule
\end{tabular}
\end{table}

\subsection{Results}
A very interesting  observation is that across almost all experiments, using the SVM on the extracted W2V2 features outperforms the simple mean-pooling based classification head (NN). 
Analyzing results on the podcast level, as done in \Cref{tbl:results}, reveals a wide spread of results,
with relative differences of up to 37\% for blocks,  52\% for interjections, 20\% for prolongations, 268\% for sound repetitions, and 61\% for word repetitions.

\Cref{tbl:results_sep28k_e_mlp} contrasts the results for the speaker agnostic splits (5/10-fold-cv), the average results over the leave-one-podcast out (LOPO) cross-validation(CV), SEP12k-CV results, and the three train-development-test splits described in \cref{ss:partitions}.
Results on the speaker agnostic CV splits are slightly more optimistic than the LOPO results except in one case. 
The effect is more pronounced in the experiments using the neural network classification head on top of the W2V2 model.   
Each CV result reported, is to be interpreted as the mean of five unique CV runs, each using different random seeds for splitting.
Results for 10-fold CV and 5-fold CV are very similar and vary only slightly across multiple validation runs, which is indicated by the small standard deviation of the results, and also shows that results are converging after multiple runs. 
Experimental results of prolongations are an outlier, being the only time LOPO results are greater than 5- and 10-fold CV.

Results for NN on the SEP-28k-E split are among the best for word repetitions and prolongations and are overall slightly optimistic compared to LOPO, SEP28k-T, SEP-28k-D, and SEP-12k results. 
SVMs achieve decent performance on SEP-28k-E, with overall results slightly less optimistic than the speaker agnostic CV experiments.
Compared to the SEP-12k evaluation scenario, results on SEP-28k-E are slightly better for interjections and prolongations, substantially better for word repetitions, and slightly worse for sound repetitions and blocks, but within the expected deviation. 
\begin{table}[t]
\caption{Results (F1-scores) for non-speaker exclusive cross-validation experiments, leave-one-podcast out (LOPO), 5-fold CV results on SEP-12k, and three different speaker exclusive splits from SEP-28k-E, omitting "SEP" in the interest of brevity. (Bl=block, Pro=prolongation, Wd=word repetition, In=interjection)}\label{tbl:results_sep28k_e_mlp}

\vspace{-2mm}
\centering
\begin{tabular}{r|r|c|c|c|c|c|c|c}
\toprule
                        &              & 5-fold-cv              &  10-fold-cv           &  LOPO                  & SEP-12k     & 28k-E        & 28k-T & 28k-D        \\
\midrule                                                                                                           
\multirow{5}{25pt}{SVM} & \textbf{Bl } & \textbf{0.36} (0.03)   &\textbf{0.36} (0.03)   &   0.34 (0.03)          & 0.34 (0.02) & 0.33             & 0.33      & 0.33            \\
                        & \textbf{In } & \textbf{0.71} (0.01)   &\textbf{0.71} (0.01)   &   0.66 (0.08)          & 0.64 (0.03) & 0.68             & 0.70      & 0.70            \\
                        & \textbf{Pro} &  0.46 (0.02)           & 0.47 (0.02)           &   \textbf{0.48} (0.04) & 0.44 (0.03) & 0.46             & 0.43      & 0.44            \\
                        & \textbf{Snd} &  0.46 (0.05)           &\textbf{0.47} (0.03)   &   0.41 (0.14)          & 0.41 (0.06) & 0.39             & 0.41      & 0.42            \\
                        & \textbf{Wd } &  0.51 (0.03)           & \textbf{0.52} (0.03)  &   0.41 (0.07)          & 0.42 (0.05) & 0.51             & 0.45      & 0.45            \\
\midrule                                                                                                              
\multirow{5}{25pt}{NN} & \textbf{Bl } &  \textbf{0.22} (0.02)   & \textbf{0.22} (0.02)  & 0.19 (0.03)            & 0.22 (0.05) & 0.19             &   0.19    & 0.21            \\
                       & \textbf{In } &  \textbf{0.69} (0.01)   & \textbf{0.69} (0.02)  & 0.62 (0.09)            & 0.66 (0.02) & 0.65             &   0.69    & 0.70            \\
                       & \textbf{Pro} &  \textbf{0.41} (0.02)   & \textbf{0.41} (0.02)  & 0.40 (0.05)            & 0.40 (0.04) & \textbf{0.41}    &   0.39    & 0.39            \\
                       & \textbf{Snd} &  \textbf{0.43} (0.02)   & 0.42 (0.03)           & 0.36 (0.14)            & 0.38 (0.03) & 0.39             &   0.37    & 0.38            \\
                       & \textbf{Wd } &  0.44 (0.02)            & 0.44 (0.03)           & 0.35 (0.07)            & 0.40 (0.01) & \textbf{0.46}    &   0.39    & 0.40            \\
\bottomrule
\end{tabular}
\end{table}

\subsection{Discussion}
It is hard to directly compare the results of all splits and strategies introduced in this paper, as they vary greatly w.r.t. the amount of training data and their speaker composition. 
Results from the experiments utilizing the LOPO strategy reveal that evaluation results on unique splits can vary considerably, and there is potential for cherry-picking particularly favourable subsets for reporting results. 
There are many possible splits and there will never be only one perfect evaluation strategy.
Despite this, common ground has to be established; otherwise, it is not possible to compare methods.
For a fair evaluation that makes sure that systems detect dysfluencies reliably, independent of the speaker, it is paramount to avoid having speech samples of speakers in the test set that are either in the development or training set. 
In scenarios such as long-term monitoring, it might be advisable to adapt models to a speaker, but this requires different methods, datasets, and evaluation scenarios. 
This is supported by the slightly optimistic results on the speaker agnostic CV splits. 

Keeping in mind what gets evaluated using a certain data split or evaluation method is important when comparing the results of different methods for dysfluency detection.
The 5-fold CV experiments on the SEP-12k subset, consisting of speech from many speakers, consistently yield lower F1-scores across most experiments. 
Even though less training data was used for training and testing, these experiments indicate that speaker-independent splits can lead to worse results.
SVMs can learn solid decision boundaries from only a few good representatives, which at least makes the SVM results a fair comparison. 

The problem of the speaker imbalance in the dataset cannot be solved for the whole dataset, no matter which speaker-exclusive split is used, as the dominant speakers, almost certainly bias the training process, no matter in which data partition they are. 
An ideal speaker-independent evaluation would be a leave-one-speaker out evaluation, but it is not feasible on large datasets and would probably harm overall progress in the interest of supposed maximum fairness. 

Therefore, we suggest using the SEP-28k-E split for swift and easy development with a fixed train-test-validation split.
Still, this does not guarantee perfect generalisation, as it is only trained on a few speakers, but the development- and test-set vary considerably and can hopefully provide realistic results.
When adapting the weights of a neural network pre-trained on data from a few speakers with transfer learning, we expect that even a few out-of-domain samples from multiple speakers will lead to greater generalisation.

The approach in this paper is limited w.r.t. the classification methods used, that was not specifically tailored for each dysfluency type. 
Therefore results vary greatly for different dysfluency types.
Especially those that need longer temporal context or a higher time-resolution, such as word-repetitions, and blocks suffer from the method employed. 
Still, all experiments achieve above chance-level results \cite{lea_sep-28k_2021}, making the demonstrated differences based on multiple data partitionings meaningful.
Even though we strongly believe that the semi-automatically generated speaker labels are valid, there are limitations. 
The splits created and used here did use the least restrictive clustering criteria in order to keep all training data.
Some episodes have hosts from other podcasts appearing as guests, or episodes of, e.g., StutterTalk feature former guests as hosts for some episodes. 
We tried to account for this with manual rules but are aware that this speaker separation will not be perfect.
If one seeks maximum speaker independence, the meta data and quality metrics provided allow the exclusion of such clips. 

\section{Conclusion}

SEP-28k is a very valuable resource when working on new methods and applications for dysfluency detection systems because it significantly increases the amount of training data available, but one has to be aware of peculiarities that might lead to problems. 
This paper contributed important insights into working with the SEP-28k dataset. 
We created and published SEP-28k-Extended, as an addition to one of the largest freely available resources containing labeled stuttered speech.
It provides semi-automatically generated speaker labels obtained using K-Means clustering on ECAPA speaker representations. 
The speaker labels are accompanied by quality metrics for these automatically generated speaker labels, enabling the creation of new speaker-independent splits with different levels of strictness. 
We hope to raise awareness about potential problems when working with SEP-28k and provide potential remedies to a few.

Based on the generated and retrieved metadata, we suggest five different possible splits, each with a different evaluation goal alongside baseline results.
Thus, establishing common ground for the future evaluation of dysfluency detection methods using the SEP-28k dataset.

\bibliographystyle{splncs04}
\footnotesize{
\bibliography{references,refs}
}
%
%
%
%

\end{document}